# High-Throughput Computational Exploration of MOFs for Short-Chain PFAS Removal

Mengru Zhang[a], Satyanarayana Bonakala[a], Taku Watanabe[b], Karim Hamzaoui[b], Guillaume Maurin[a,c,*]

[a]*ICGM, Univ. Montpellier, CNRS, ENSCM, Montpellier, France*

[b]*Matlantis Corporation, Otemachi, Chiyoda-ku, 100-0004 Tokyo, Japan*

[c]*Institut Universitaire de France, France*

**Abstract**

Short-chain per- and polyfluoroalkyl substances (PFASs) are increasingly replacing regulated long-chain PFASs, yet they remain challenging to remove from water due to their high persistence, mobility, and weak affinity toward conventional adsorbents. In this work, we developed a hybrid high-throughput computational screening (HTCS) strategy to identify high-performance MOFs for the selective adsorption of perfluorobutanoic acid (PFBA), a representative short-chain PFAS, from water. The workflow begins with a curated MOF dataset and employs Monte Carlo (MC) simulations based on synergistic use of a classical universal force field (UFF) and a universal machine-learned interatomic potential (u-MLIP), enabling scalable and quantitatively accurate prediction of adsorption across large MOF databases. A set of promising MOFs initially identified using UFF-based HTCS, that combine strong PFBA affinity and high PFBA/$H_2O$ selectivity were re-evaluated with u-MLIP to refine adsorption predictions and to assess guest-induced framework flexibility, enabling the exclusion of materials with unfavourable high water–framework interactions. Ultimately, four high-performance MOFs were identified that optimally balance strong PFBA interactions, high PFBA selectivity over water, and practical considerations including sustainability, water stability, and synthetic feasibility. This study demonstrates that combining classical force fields with u-MLIPs enables scalable, quantitatively accurate MOF adsorption screening and establishes transferable principles for the rational design of adsorbents targeting short-chain PFAS.

*Keywords:* High-throughput computational screening, MOFs, Short-chain PFAS removal, machine learned interatomic potential, structure–performance relationship

## 1. Introduction

Per- and polyfluoroalkyl substances (PFASs) are a broad class of synthetic fluorinated organic compounds that have been widely used in industrial and commercial applications.[1–3] Owing to their high chemical stability and resistance to degradation, PFASs are persistent in the environment, leading to a widespread contamination of water and soil.[4] Human exposure to PFASs and their accumulation in biological systems have been associated with various adverse health effects, including immune system suppression, metabolic dysfunction, and increased risks of certain cancers.[5] Accordingly, the production and use of long-chain PFASs, typically defined as compounds containing six or more perfluorinated carbon atoms, have been therefore increasingly regulated or banned in many countries.[6] This regulatory shift has led to the introduction and widespread use of short-chain PFASs as alternatives.[7] Although short-chain PFASs bioaccumulate less than their long-chain counterparts, they are far from benign: their extreme persistence and high mobility in aquatic environments enable widespread dispersion, leading to sustained ecological exposure and potential human health risks. Despite this growing concern, most PFASs remediation strategies have focused on long-chain species, leaving the adsorptive removal of short-chain PFASs largely unexplored.[8,9] Moreover, conventional adsorbents, such as activated carbons, ion-exchange resins, polymers and zeolites, have shown limited effectiveness toward short-chain PFASs, further highlighting the need for alternative materials combined with complementary separation strategies.[10–12]

Among emerging porous materials for challenging aqueous separations, metal–organic frameworks (MOFs) have attracted growing attention as promising adsorbents.[13,14] MOFs are crystalline nanoporous materials constructed from metal ions or metal-based clusters coordinated with organic linkers, forming extended two- or three-dimensional frameworks with well-defined pore architectures.[13] Their modular construction provides exceptional structural diversity, allowing precise control over pore size, topology, and internal surface chemistry. This tunability enables the design of tailored pore environments, accessible coordination sites, and adjustable polarity, features that are particularly advantageous for the selective capture of persistent pollutants such PFASs.[13–16] To date, numerous MOFs have been identified as efficient adsorbents for PFASs removal from water.[17] Notable examples include MOFs from the MIL (Materials of Institut Lavoisier) series [18–21], ZIF (Zeolitic Imidazolate Framework) families[22], UiO-66 derivatives[23–27] and others[19,28,29]. However, these studies have primarily focused on adsorption of long-chain PFAS, and only a limited number of MOFs, such as UiO-66 and its derivatives,[23–25] NU-1000,[29] and MOF-808,[30] have been evaluated for short-chain PFASs adsorption (e.g., PFBA, PFPeA, PFHxA, and PFBS). Overall, these previous studies reported a reduced effectiveness of the MOFs toward short-chain PFASs, with adsorption performance consistently declining as the PFAS carbon chain shortens. Therefore, we are far from unlocking the immense potential of MOFs for PFASs capture in particular because their limited exploration has so far been driven entirely on serendipity.

To bridge this critical gap, we implemented a hybrid high-throughput computational screening (HTCS) strategy merging classical generic universal force field (u-FF) and universal machine learned interatomic potential (u-MLIP)[31,32] to identify MOFs with high potential for the selective removal of Perfluorobutanoic Acid (PFBA), a representative short-chain PFAS, from water. Our screening leverages a curated database of computation-ready MOFs and follows a stepwise multi-level narrowing process, as illustrated in Fig. 1. The workflow integrates successive screening criteria that reflect both adsorption performance and practical considerations. Initial structural pre-screening based on pore-limiting diameter (PLD) is completed by thermodynamic evaluations of $H_2O$ and PFBA affinity along ideal PFBA/$H_2O$ selectivity quantified via the Henry's law constant of PFBA ($K_{H,PFBA}$) and water ($K_{H,H_2O}$), and the PFBA-to-water Henry's law constants ratio ($S_{ads, PFBA/H_2O}$) respectively, obtained from Widom's test particle insertion Monte Carlo simulations. Sustainability, water stability and synthetic feasibility criteria are then applied to further narrow down the list of promising candidates. The performance of these top MOF candidates is further refined using u-MLIP, enabling an accurate description of MOF–guest interactions and MOF framework flexibility. In parallel, a quantitative structure–performance relationship (QSPR) analysis reveals how key structural features govern PFBA adsorption affinity and selectivity in MOFs. Through this multi-level screening, rationalization and validation workflow, we ultimately identify the most promising adsorbents for selective PFBA capture from water.

## 2. Computational Methods

### 2.1. MOFs structure database

We considered a curated MOF database comprising 18,559 structurally consistent and computational-ready structures reported earlier.[31] The geometric features of all MOFs, including pore-limiting diameter (PLD), largest cavity diameter (LCD), accessible surface area (ASA), accessible Volume ($V_{acc}$) and void fraction, were computed using Zeo++.[33]

### 2.2. PFBA and water models

For classic force field simulations, the PFBA molecule was modelled by the charged Lennard–Jones (LJ) site models derived from the united-atom TraPPE and OPLS-AA force fields.[34] PFBA was treated as a rigid molecule, with the perfluoroalkyl chain ($-CF_3$ and $-CF_2$ groups) represented by united-atom sites and the carboxyl group modeled explicitly (Fig. S1 and Table S1). Two PFBA conformations, linear and boat-type, were initially considered. Density Functional Theory (DFT)-based geometry optimizations demonstrated that the linear PFBA conformation is energetically more favourable than the boat-type geometry (Fig. S2), and was therefore adopted in all MC simulations. Water molecules were described using the well-established TIP4P-Ew model.[35]

### 2.3. Classical Widom's test particle insertion Monte Carlo Simulations

Because PFBA removal is performed in aqueous environments, competitive adsorption by water represents a critical limiting factor. We quantified the intrinsic affinity of MOFs for water using the Henry's law constant $K_{H,H_2O}$, which directly probes the water–framework interaction strength in the infinite-dilution limit. These calculations were performed by Widom's test particle insertion Monte Carlo (MC) method at 298 K. A threshold value of $K_{H,H_2O}$ was therefore imposed to filter out materials with strong water affinity, ensuring that the remaining candidates exhibit adsorption behavior mostly governed by preferential PFBA–framework interactions.

To assess PFBA affinity, we next evaluated the enthalpy of adsorption at infinite dilution, $\Delta H^0_{ads,PFBA}$, as well as the Henry's law constant of PFBA, $K_{H,PFBA}$, both quantities being equally calculated using Widom's test particle insertion MC method. Finally, to characterize the selectivity of the filtered MOFs for adsorbing PFBA from water, we used the ideal adsorption selectivity metric: $S_{ads, PFBA/H_2O} = K_{H,PFBA}/K_{H,H_2O}$, where $K_{H,PFBA}$ and $K_{H,H_2O}$ denote the Henry's law constants of PFBA and water, respectively. All classical-FF MC simulations were carried out using the RASPA 2.0 simulation package.[36] Each simulation consisted of 500,000 Monte Carlo steps to ensure statistical convergence. MOF frameworks were treated as rigid during the simulations. Non-bonded interactions were described using a combination of LJ and Coulombic terms. LJ parameters for MOF atoms were assigned from the UFF force field,[37] and cross-interaction parameters were determined using the Lorentz–Berthelot mixing rules.[38] Partial charges for MOF atoms were assigned using the MEPO-ML (MOF Electrostatic POtential–Machine Learned) model.[39] Long-range electrostatic interactions were evaluated using the Ewald summation. Periodic boundary conditions were applied in all three dimensions. A cut-off distance of 12.0 Å was used for both the LJ potential and the real-space component of the Ewald sum. Simulation cells of the MOFs were constructed to ensure that the shortest box length exceeds twice this cut-off distance.

*2.4. Force Field prediction validated by u-MLIP and DFT calculations*

*2.4.1. u-MLIP-based simulations*

To refine the adsorption metrics predicted for the six top-performing MOFs identified from UFF-based HTCS, they were re-evaluated using the PreFerred Potential (PFP) u-MLIP (version 8.0.0).[32,40] Using this approach, $K_H$ and $\Delta H^0_{ads}$ for both H$_2$O and PFBA were recalculated in each selected MOF. Dispersion interactions were explicitly accounted for through the Perdew–Burke–Ernzerhof (PBE) together with Grimme's D3 dispersion correction.[41–43] Each simulation consisted of 100,000 MC cycles to ensure statistical convergence.

To further assess the effect of adsorption-induced MOF framework flexibility on PFBA affinity, we repeated the calculation of the adsorption metrics ($K_H$ and $\Delta H^0_{ads}$) for the same six MOFs using u-MLIP–fully geometry optimized PFBA-loaded MOF structures (PFBA@MOF). Both atomic positions and unit cell parameters of the PFBA@MOF structures containing one-PFBA-per-pore were relaxed using the u-MLIP with the BFGS optimizer while imposing a force convergence threshold of 0.005 eV Å$^{-1}$. The PFBA molecules were further removed from these relaxed MOF structures, and Widom's test particle insertion MC simulations were applied to compute $K_H$ and $\Delta H^0_{ads}$ for PFBA using u-MLIP. The same protocol was applied for H$_2$O. All these calculations were performed on the MATLANTIS platform (academic license), where the u-MLIP is accessed through the Estimator module coupled to an ASECalculator for energy and force evaluations within the Atomic Simulation Environment (ASE).[44]

*2.4.2. u-MLIP benchmarking against DFT calculations*

To assess the reliability of the u-MLIP for an accurate description of MOF–PFBA interactions, the u-MLIP was benchmarked against DFT calculations. The 6 preliminary identified PFBA@MOF structures were fully geometry at the DFT level using the Quickstep module in the CP2K package.[45] The PBE exchange–correlation functional was employed together with Grimme's D3 dispersion correction.[41–43] Core electrons were described using Goedecker–Tetter–Hutter (GTH) pseudopotentials,[46–48] and valence electrons were expanded in the Gaussian and plane-wave (GPW) framework using triple-ζ valence MOLOPT basis sets with a plane-wave density cutoff of 600 Ry, as implemented in CP2K/Quickstep (version 2024.1). The resulting adsorption energy ($E_{ads}$ =E(PFBA@MOF)-E(MOF)-E(PFBA) with $E$(PFBA@MOF), the total energy of the geometry optimized MOF‑PFBA configurations, while $E$(MOF) and $E$(PFBA) are the single-point energies of (i) the MOF framework for which PFBA is removed and (ii) the isolated PFBA molecule respectively. Adsorption-induced volume change ($V_{change}$%) derived by these DFT-calculations were compared to the data obtained at the u-MLIP level.

## 3. Results and Discussion

Fig. 1 outlines the HTCS workflow implemented in this study. The strategy comprises three successive stages: (i) pore-geometry-based pre-screening to ensure accessibility of PFBA within the pores of the selected MOFs, (ii) water–framework affinity filtering to suppress MOFs with highly competitive water adsorption, (iii) UFF-guided identification of MOFs candidates with potential for selective PFBA adsorption through a subsequent evaluation of PFBA/H$_2$O selectivity and (iv) refinement of PFBA adsorption performance using u-MLIP to accurately capture MOF–guest interactions and framework flexibility. Each stage is detailed in the following sections.

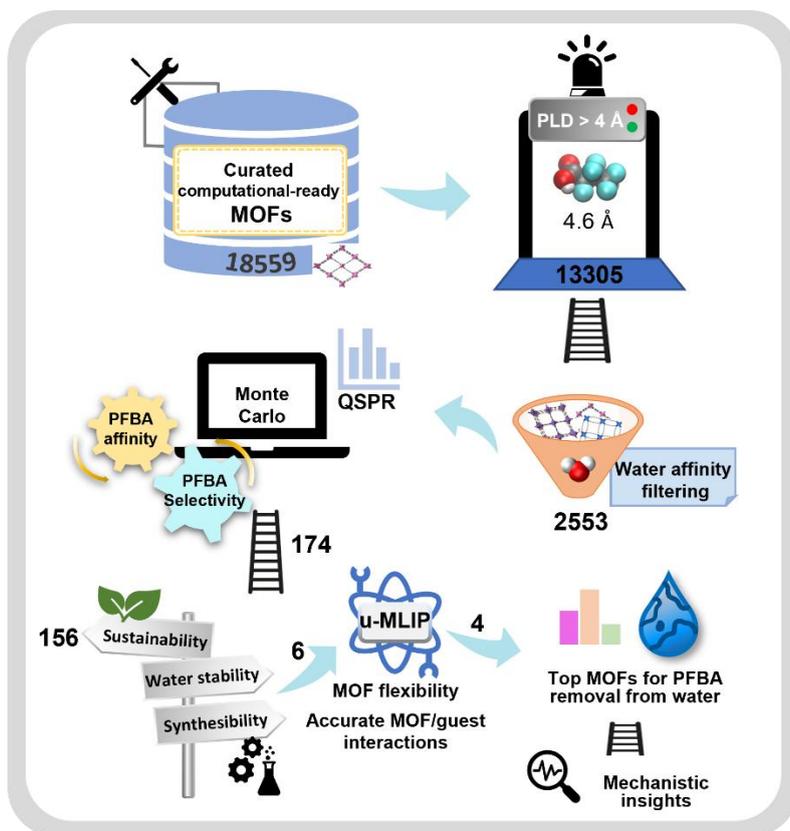

Fig. 1. High-throughput computational screening workflow used to identify promising MOFs for PFBA adsorption towards rationalization.

*3.1. UFF-based HTCS screening of MOFs for PFBA selective adsorption towards rationalization*

The computational-ready MOFs in our curated database (18,559 entries) were initially screened by excluding structures with PLDs below 4 Å, as such pores are too small to accommodate PFBA, which has a kinetic diameter of 4.6 Å. This filtering resulted in a selection of 13,305 MOFs. Following this geometric prescreening, a water–framework affinity filtering was applied to exclude MOFs prone to strong competitive water adsorption. In the context of aqueous adsorption, we calculated $K_{H,H_2O}$ for all these MOFs using the Widom's test particle insertion MC method, and we imposed a threshold value of $K_{H,H_2O} = 10^{-5}$ mol kg$^{-1}$ Pa$^{-1}$ to enable the exclusion of MOF frameworks in which strong water–framework interactions would lead to extensive adsorption site solvation and competitive displacement of PFBA. This criterion therefore enables the selection of MOFs in which water acts predominantly as a solvent rather than an adsorbate, preserving accessible adsorption sites for selective PFBA uptake. This yields to a subset of 2553 MOFs potentially suitable for PFBA removal. Using the same Widom's test particle insertion MC method, the PFBA affinity of this MOF subset was evaluated by calculating the enthalpy of adsorption at infinite dilution $\Delta H^0_{ads,PFBA}$ and their potential selective PFBA adsorption over water was assessed by calculating their ideal adsorption selectivity values $S_{ads, PFBA/H_2O}$. These two calculated adsorption metrics are summarized in Figure 2a. More negative PFBA adsorption enthalpies at infinite dilution ($\Delta H^0_{ads,PFBA}$) and higher adsorption selectivity $S_{ads, PFBA/H_2O}$, reflecting strong PFBA affinity and effective competition against water, respectively, are therefore prioritized to identify the most promising MOFs for the targeted application.

These overall adsorption data were rationalized to assess the impact of the MOF structural factors on their adsorption performances. $S_{ads, PFBA/H_2O}$, $K_{H, PFBA}$ and $\Delta H^0_{ads,PFBA}$ were analyzed as functions of key geometric and textural MOF descriptors, including PLD, LCD, void fraction, framework density ($\rho$), ASA and $V_{acc}$. The corresponding Pearson correlation coefficients quantifying these structure–performance relationships are summarized in the heat map shown in Fig. 2b. $\Delta H^0_{ads,PFBA}$ shows very strong positive correlations with log($S$) (Pearson correlation coefficient r = 0.91), indicating that strong PFBA–framework interactions directly translate into high PFBA/H$_2$O selectivity. In the mean-time, $\Delta H^0_{ads,PFBA}$ is negatively correlated with key geometric descriptors, most notably void fraction (r = –0.67), ASA (r = –0.61), LCD (r = –0.59) and PLD (r = –0.49). These trends indicate that PFBA adsorption is favored in more confined pore environments. Consistent behavior is observed for $\log(K_{H,PFBA})$ and $\log S$ (PFBA/H$_2$O), which both increase with decreasing void fraction (r = –0.54 and –0.59, respectively) and ASA (r = –0.48 and –0.55 respectively). This finding highlights that high porosity or large surface area alone does not ensure effective PFBA adsorption; instead, excessively open pore structures weaken host–guest interactions and diminish adsorption performance. By contrast, $\rho$ correlates positively with PFBA adsorption affinity (r = 0.48 with $-\Delta H^0_{ads,PFBA}$), suggesting that more compact frameworks generally provide stronger interacting environments. To sum up, the heat map analysis demonstrates that PFBA affinity and PFBA/H$_2$O selectivity are primarily governed by pore confinement rather than overall porosity. MOFs with moderately small pore sizes and limited void space are therefore expected to offer optimal performance by strengthening PFBA–framework interactions.

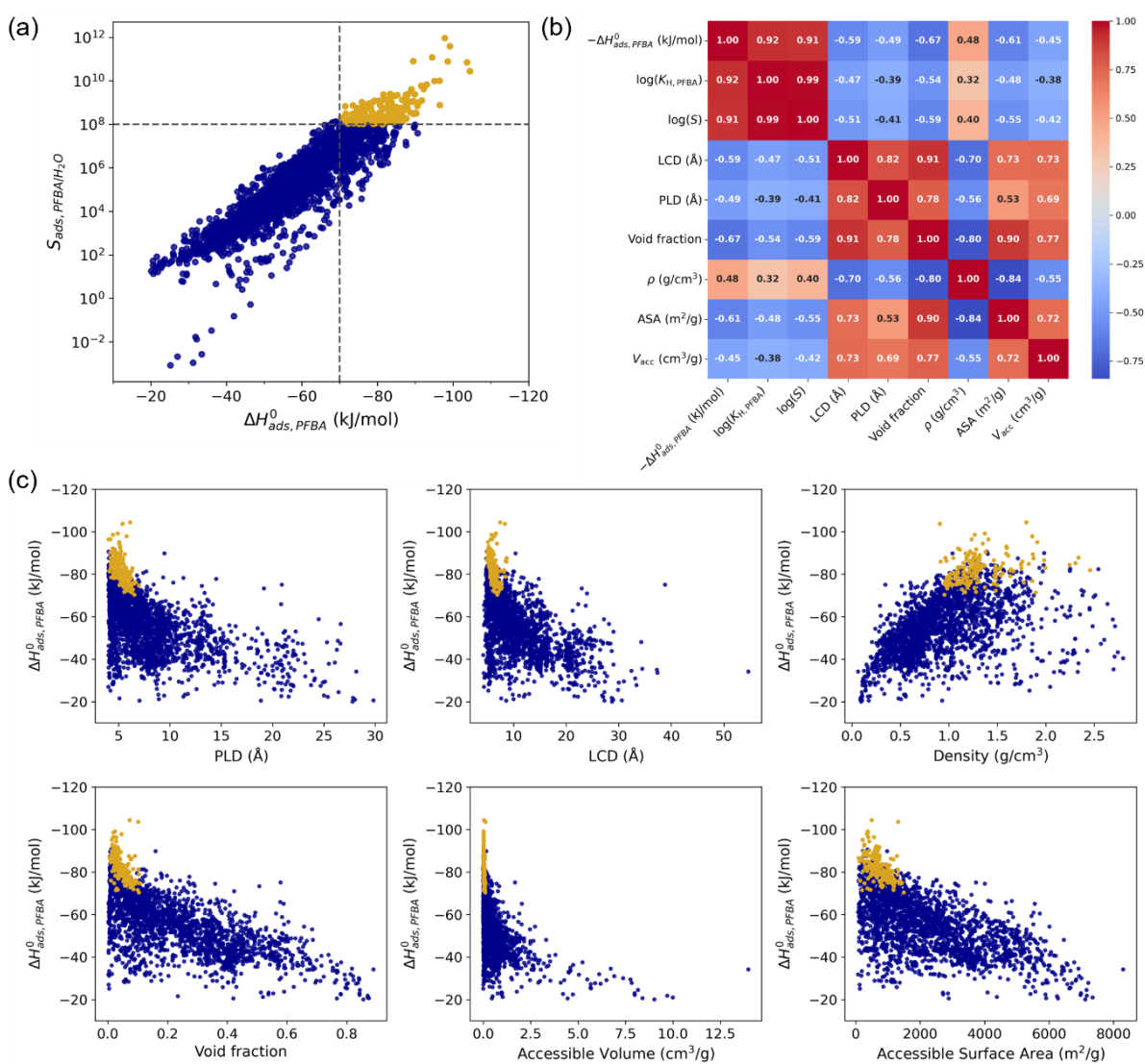

Fig. 2. Correlation analysis between PFBA adsorption performance and structural descriptors for the 2553 screened hydrophobic MOFs. (a) Ideal adsorption selectivity, $S_{ads, PFBA/H_2O}$, as a function of the PFBA adsorption enthalpy at infinite dilution, $\Delta H^0_{ads,PFBA}$. The dashed lines indicate the screening thresholds of $\Delta H^0_{ads,PFBA} < -70$ kJ mol$^{-1}$ and $S_{ads, PFBA/H_2O} > 10^8$, identifying 174 high-performing candidates. (b) Pearson correlation heat map for the 2553 screened MOFs relating PFBA adsorption thermodynamics ($-\Delta H^0_{ads,PFBA}$, $\log(K_{H,PFBA})$, and $\log(S_{ads, PFBA/H_2O})$) to key structural descriptors (PLD, LCD, void fraction, framework density $\rho$, accessible surface area, and accessible volume). (c) Scatter plots of $\Delta H^0_{ads,PFBA}$ versus the structural and textural MOF descriptors

## 3.2. UFF-guided identification of MOFs for PFBA removal

To prioritize materials with both strong PFBA affinity and high selectivity towards water adsorption, we considered two criteria: $\Delta H^0_{ads,PFBA} < -70$ kJ mol$^{-1}$ and $S_{ads, PFBA/H_2O} > 10^8$ that delimits a zone in dark yellow in Fig. 2(a). Application of these thresholds yields a subset of 174 MOFs with attractive predicted adsorption performance. To further examine the structural characteristics of top-performing materials, the distribution of these 174 MOFs across key structural and textural descriptors are highlighted in Fig. 2c. Compared to the full set of 2553 MOFs, the selected candidates are predominantly located in regions corresponding to smaller pore sizes, lower void fractions, and higher framework densities. This observation is consistent with the correlation analysis discussed above and confirms that selective PFBA adsorption preferentially occurs in moderately confined pore environments.

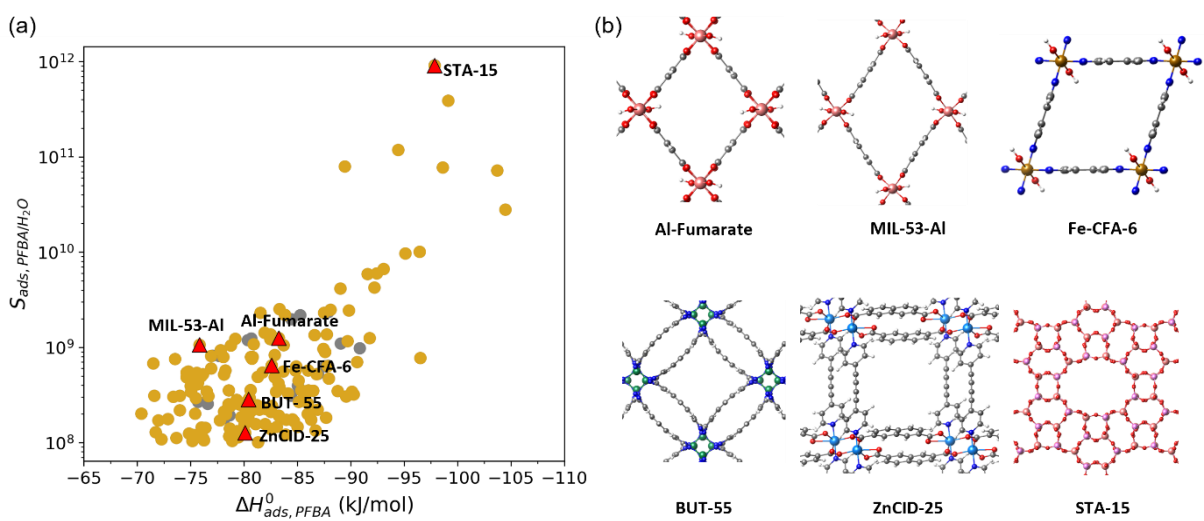

Fig. 3. (a) Ideal adsorption selectivity, $S_{ads, PFBA/H_2O}$, as a function of the PFBA adsorption enthalpy at infinite dilution $\Delta H^0_{ads,PFBA}$ for the 156 out of 174 best MOF candidates identified in Figure 2a after excluding structures containing rare-earth metals based on sustainability considerations. (b) Final selection of six top-performing candidates, experimentally feasible, including five MOFs and one aluminophosphate, chosen for detailed validation and further investigation. Element colour scheme: C: grey; H: white; O: red; N: dark blue; P: purple; Fe: Gold; Zn: blue; Al: pink; Co: green.

Beyond adsorption metrics, practical considerations related to material sustainability and feasibility were incorporated. MOFs containing rare-earth metals were excluded due to their limited abundance, high environmental concerns and costs, resulting in 156 remaining candidates, as summarized in Fig. 3(a). These materials therefore represent a more realistic pool of high-performance adsorbents composed of commonly available elements. In addition, the synthetic feasibility and water stability of the remaining structures were checked based on the reported literature.[49–54] From this perspective, Al-fumarate, MIL-53-Al, BUT-55, STA-15, and ZnCID-25 were shown to be water stable, while Fe-CFA-6 belongs to the family of high-valent metal-azolate MOFs known to generally exhibit high resistance to water/hydrolysis. Through this process, six top candidates, five MOFs and one aluminophosphate were selected for detailed investigation, as illustrated in Fig. 3(b). Their key pore descriptors (e.g., LCD, PLD, Density, void fraction, Vacc, and ASA) are summarized in Table S2 (ESI). These most promising materials were selected for subsequent high-level validation and performance in-depth analysis.

## 3.3. Refining promising candidates using u-MLIP: impact of adsorption-induced relaxation

Beyond the use of generic classical force fields, which may only provide a qualitative description of host–guest interactions, the preliminary UFF-based HTCS treats MOFs as rigid frameworks. While this approach enables efficient exploration of large structural spaces, it inherently neglects adsorption-induced structural responses and their potential impact on the MOF adsorption performances. To address these two limitations, the top-performing candidates identified from the UFF-based screening were further examined using u-MLIP.

Table 1 PFBA adsorption energies ($E_{ads}$) and PFBA-adsorption induced volume changes ($\Delta V\%$) obtained by u-MLIP and DFT calculations for the six selected MOFs. Deviation of the adsorption energies between u-MLIP and DFT is equally reported.

| MOFs | $E_{ads,\text{u-MLIP}}$(kJ/mol) | $E_{ads,\text{DFT}}$(kJ/mol) | Deviation ($E_{ads}$ u-MLIP vs DFT) | $\Delta V\%(mlip)$ | $\Delta V\%(dft)$ |
|---|---|---|---|---|---|
| STA-15 | -89.8 | -97.6 | 8.7 | -1.2 | -2.3 |
| Al-fumarate | -89.3 | -95.3 | 6.7 | 1.0 | -0.8 |
| Fe-CFA-6 | -121.9 | -115.1 | 5.6 | -6.9 | -6.2 |
| BUT-55 | -82.0 | -89.2 | 8.8 | -6.3 | -8.3 |
| ZnCID-25 | -83.1 | -90.7 | 9.1 | -2.6 | -9.1 |
| MIL-53-Al | -86.5 | -93.7 | 8.3 | -5.4 | -7.9 |

The reliability of the u-MLIP for modeling PFBA adsorption in the 6 top-performers adsorbents was firstly validated against DFT calculations. Adsorption energies ($E_{ads}$) and adsorption-induced MOF volume changes ($\Delta V\%$) were evaluated for MOF structures containing one PFBA molecule per pore, with the results summarized in Table 1. Overall, the u-MLIP reproduces the DFT adsorption energies with good fidelity across all six MOFs with relative deviations in the range 5–9%. This indicates that the u-MLIP reliably captures PFBA–framework interactions across chemically and structurally diverse systems. In parallel, both u-MLIP and DFT predict consistent adsorption-induced structural changes of these MOF frameworks. In all cases, the absolute volume change upon PFBA adsorption remains below 10% ($|\Delta V\%| < 10\%$), suggesting that the PFBA adsorption induces only modest framework relaxation without significant structure contraction or expansion for this series of explored MOFs. Combined with its previously validated performance for water–MOF interactions,[31] the selected u-MLIP provides a robust basis for refining the adsorption performance predicted for the best adsorbent candidates.

Table 2
$\Delta H^0_{ads,PFBA}$ and $S_{ads,\ PFBA/H_2O}$ for the 6-selected MOFs calculated using rigid classical force-field models and u-MLIP implementing adsorption-induced framework relaxation.

| | FF | | u-MLIP | | | |
|---|---|---|---|---|---|---|
| MOFs | $\Delta H^0_{ads,PFBA}$ | $S_{ads}$ | $\Delta H^0_{ads,PFBA}$ | $\Delta H^0_{ads,H_2O}$ | $\Delta V\%$ ($H_2O$) | $S_{ads}$ |
| Fe-CFA-6 | -82.6 | 6.5E+08 | -78.5 | -39.0 | -8.3 | 1.3E+05 |
| ZnCID-25 | -80.1 | 1.3E+08 | -65.4 | -35.0 | 2.9 | 1.7E+04 |
| BUT-55 | -80.6 | 2.8E+08 | -65.2 | -27.7 | -1.0 | 9.5E+03 |
| Al-fumarate | -83.2 | 1.3E+09 | -66.6 | -33.9 | -4.6 | 7.1E+03 |
| MIL-53-Al | -75.8 | 1.1E+09 | -63.3 | -51.3 | -33.8 | 6.0E+02 |
| STA-15 | -97.8 | 9.1E+11 | -89.9 | -87.8 | -0.7 | 4.6E-03 |

Building on the validated u-MLIP framework, predictions of the adsorption metrics for the top adsorbents were refined. As summarized in Table 2, $\Delta H^0_{ads,PFBA}$ predicted by u-MLIP considering a full flexibility of the MOF frameworks are consistent with those obtained from UFF in terms of energetics sequence although showing significant deviations up to ~17 kJ mol$^{-1}$. These results reveal that PFBA adsorption induces only limited structural relaxation, such that UFF remains sufficient to provide a reasonable first-order evaluation of PFBA affinity for the selected adsorbents. In contrast, a markedly different behavior is observed for water adsorption. MIL-53-Al is shown to exhibit a substantial volume contraction upon H$_2$O (-33.8%) as reported earlier experimentally[55,56] while there is only a very minor cell volume contraction for these other adsorbents. Furthermore, MIL-53-Al and STA-15 exhibit substantially high u-MLIP predicted water adsorption enthalpies, with associated $\Delta H^0_{ads,H_2O}$ of -51.3 and -87.8 kJ mol$^{-1}$, respectively, which compromises their potential to selectively adsorb PFBA from water, both porous materials being therefore excluded due to their unfavorable competitive PFBA adsorption behavior over water.

After this refinement step, four MOFs remain as the most promising candidates, combining suitable pore sizes, strong PFBA affinity, high PFBA/H$_2$O selectivity, and favorable sustainability and synthesizability. The PFBA adsorption configurations obtained from PFBA adsorption–induced relaxed frameworks by u-MLIP are shown in Fig.

4, providing molecular-level insight into the dominant host–guest interactions in these 4 MOFs. In Fe-CFA-6 and Al-fumarate, PFBA adsorption is primarily driven by hydrogen bonds between the PFBA carboxyl group and framework oxygen sites, with O···H distances of 2.0 and 1.9 Å in Al-fumarate and approximately 1.9 Å in Fe-CFA-6. In contrast, adsorption in the two other candidates is dominated by confinement-driven van der Waals interactions. In BUT-55, PFBA is stabilized by close contacts with the MOF pore wall, as evidenced by O···H distance of 2.9 Å; while in ZnCID-25, PFBA exhibits short F···H distance (3.0 Å), emphasizing the complementary roles of specific hydrogen bonding and pore confinement in governing PFBA adsorption behavior.

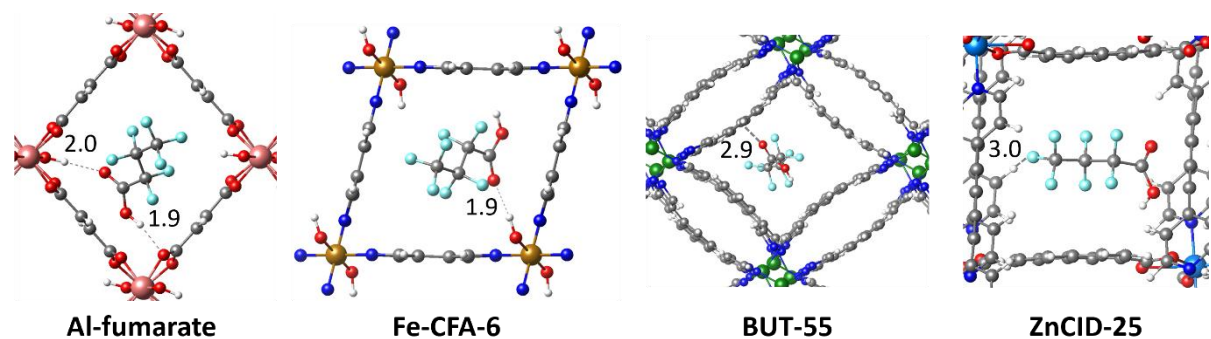

Fig. 4. Representative PFBA adsorption configurations in the four top-ranked MOF candidates obtained from PFBA adsorp- tion–induced relaxed structures simulated by u-MLIP. Numbers indicate interatomic distances (in Å). Element colour scheme: C: grey; H: white; O; red; N: dark blue; P: urple; Fe: Gold; Zn: blue; Al: pink; Co: green.

## 4. Conclusions

In this work, we developed a hybrid HTCS workflow that synergistically integrates classical force-field- and universal machine-learned interatomic potential-based Monte Carlo simulations to identify top-performing MOFs for the selective removal of PFBA from water. Starting from a curated MOF database, a hierarchical screening strategy, comprising pore-size pre-screening, water–framework affinity filtering, PFBA affinity and PFBA/$H_2O$ selectivity assessment, along with adsorption data rationalization, was employed to progressively narrow the candidate space toward high-performance materials. Structure–performance analysis revealed that pore confinement is the primary structural feature that governs strong PFBA affinity and high PFBA/$H_2O$ selectivity. Classical force-field first identified promising candidates that were further re-evaluated using u-MLIP. This refining step evidenced that PFBA adsorption induces only minor structural relaxation of their frameworks, resulting in PFBA affinities consistent with classical force-field predictions. Water was demonstrated to provoke substantial framework rearrangements and strong water–MOF interactions that significantly compromise selectivity. By explicitly accounting for these effects, our workflow eliminated candidates with high water affinity and ultimately identified four MOFs that optimally balance strong PFBA interactions, high PFBA selectivity over water, and practical considerations including sustainability, water stability, and synthetic feasibility. More broadly, this study highlights the critical importance of coupling scalable force-field screening with MLIP to achieve reliable predictions for adsorption-based separations in aqueous environments. Beyond revealing high-performance MOF sorbents for short-chain PFAS removal, the present work establishes transferable structure–performance principles and a generalizable computational workflow for the identification of advanced adsorbents targeting challenging aqueous contaminants.

**CRediT authorship contribution statement**

M.Z.: Writing – original draft, review & editing, Visualization, Validation, Formal analysis, Data curation. S.B.: Computational calculations related to u-MLIP, Validation, Data curation, Writing – review & editing. T.W.: Writing – review & editing, K.H.: Writing – review & editing. T.W. and K.H. provided support with the MATLANTIS framework. G.M.: Writing – review & editing, Supervision, Project administration, Conceptualization.

**Declaration of competing interest**

The authors declare that they have no known competing financial interests or personal relationships that could have appeared to influence the work reported in this paper.

**Data availability**

Data will be made available on request.

**Acknowledgements**

The computational work was performed using HPC resources from GENCI-CINES (Grant A0180907613) and MATLANTIS. G.M. thanks Institut Universitaire de France for the Senior Chair.

**Appendix A. Supplementary data**

Supplementary data to this article can be found online.

# High-throughput computational exploration of MOFs for short-chain PFAS removal

Mengru Zhang[a], Satyanarayana Bonakala[a], Taku Watanabe[b], Karim Hamzaoui[b], Guillaume Maurin[a,c,*]

[a]*ICGM, Univ. Montpellier, CNRS, ENSCM, Montpellier, 34095, France*
[b]*Matlantis Corporation, Otemachi, Chiyoda-ku, 100-0004 Tokyo, Japan*
[c]*Institut Universitaire de France, France*

**Corresponding Author**

*E-mail addresses: guillaume.maurin1@umontpellier.fr

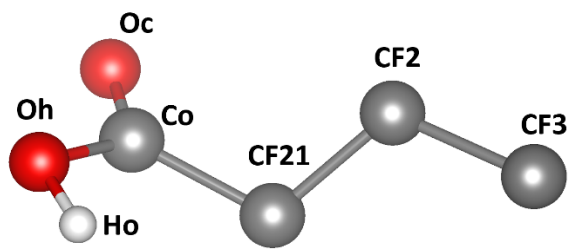

Fig. S1. The united-atom model of PFBA with atomic labels after optimization.

Table S1
Atom types and force field parameters for PFBA and water model.

| Molecule | Atom types | ε (kJ/mol) | σ (nm) | Charge (e) |
|---|---|---|---|---|
| PFBA | CF3 | 0.723 | 0.436 | 0 |
|  | CF2 | 0.229 | 0.473 | 0 |
|  | CF21 | 0.229 | 0.473 | 0.12 |
|  | Co | 0.341 | 0.390 | 0.42 |
|  | Oc | 0.657 | 0.305 | -0.45 |
|  | Oh | 0.773 | 0.302 | -0.46 |
|  | Ho | 0.000 | 0.000 | 0.37 |
| $H_2O$ (TiP4P-Ew) | Ow | 0.681 | 0.316 | 0 |
|  | Hw | 0 | 0 | 0.52422 |
|  | M | 0 | 0 | −1.04844 |

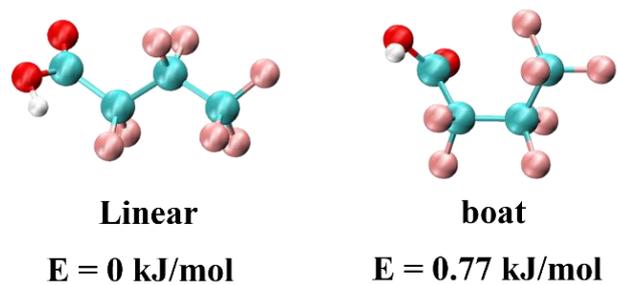

Linear  boat
E = 0 kJ/mol  E = 0.77 kJ/mol

Fig. S2. DFT-optimized PFBA conformations (linear and boat model); the linear conformer is more stable and was used in this work. Colour scheme: O, red; C, turquoise; F, pink.

Table S2
Key structural and textural descriptors of the 6 top performing MOFs.

| MOF | LCD (Å) | PLD (Å) | Density (g/cm$^3$) | Void fraction | $V_{acc}$ (cm$^3$/g) | ASA (m$^2$/g) |
|---|---|---|---|---|---|---|
| Fe-CFA-6 | 5.62 | 4.98 | 1.26 | 0.052 | 0.041 | 975 |
| Al-Fumarate | 5.46 | 5.02 | 1.10 | 0.050 | 0.045 | 1050 |
| ZnCID-25 | 6.80 | 6.12 | 1.08 | 0.072 | 0.066 | 920 |
| MIL-53-Al | 6.46 | 6.19 | 1.01 | 0.094 | 0.094 | 1349 |
| STA-15 | 6.31 | 5.77 | 1.86 | 0.046 | 0.025 | 343 |
| BUT-55 | 5.75 | 5.45 | 1.07 | 0.056 | 0.052 | 1144 |